\documentclass[a4paper,11pt]{article}

\usepackage{jheppub} 

\usepackage[T1]{fontenc} 

\title{\boldmath Electroweak-Skyrmion as Topological Dark Matter}

\author[a,b]{Ryuichiro Kitano}
\author[a]{and Masafumi Kurachi}
\affiliation[a]{Theory Center, High Energy Accelerator Research Organization (KEK), Tsukuba 305-0801, Japan}
\affiliation[b]{The Graduate University for Advanced Studies (Sokendai), Tsukuba 305-0801, Japan}

\emailAdd{Ryuichiro.Kitano@kek.jp}
\emailAdd{Masafumi.Kurachi@kek.jp}

\abstract{We show the existence of a nontrivial topological configuration of the Higgs field in the Standard Model with the Skyrme term. It is shown that the current upper bound of the mass of the topological object is about 34 TeV. We discuss the impact of the existence of the topological object on cosmology.}

\preprint{KEK-TH-1908}

\begin{document} 
\maketitle
\flushbottom

\section{Introduction}
The nature of the dark matter remains a mystery of the Universe today. Numbers of candidates for the dark matter have been proposed based on models beyond the Standard Model (SM). Here, we show the existence of a dark matter candidate in the SM with a minimal addition of a higher dimensional operator. The idea is analogous to that of the Skyrme model~\cite{Skyrme:1961vq}, in which nucleons are described by nontrivial toplogical configurations of the Nambu-Goldstone (NG)-boson fields  (pions) associated with the chiral symmetry breaking triggered by the quantum chromodynamics (QCD). Since the structure of the global symmetry breaking in the electroweak symmetry breaking (EWSB) sector of the SM is exactly the same as that of the chiral symmetry breaking in QCD, the argument by Skyrme in the QCD chiral Lagrangian is directly applied to the electroweak (EW) chiral Lagrangian by a simple scale-up of the relevant energy scale. However, there is one important difference between the low-energy physics of QCD and the EWSB sector of the SM: that is the existence of a light scalar particle, the Higgs boson. Actually, from the generalized analysis of the Skyrme model, it is known that the existence of a scalar resonance would generate the operator which destabilize the Skyrmion configuration when it is integrated out from the theory~\cite{Ellis:2012bz,Ellis:2012cs}. However, that does not immediately mean the non-existence of  topological objects in the SM. 
In order to clarify the (non-)existence of such an object, it is necessary to study the system with the scalar field included as a dynamical degree of freedom. In this paper, we derive the equations which should be satisfied by the topological configuration in a coupled system of the NG and the Higgs fields, and show that a non-trivial topological configuration actually exists in the system. The mass of the topological object, which we call the EW-Skyrmion, is related to the magnitude of the coefficient of the $O(p^4)$ term in the Lagrangian. Therefore the experimental bound on that coefficient places the upper bound on the mass of the EW-Skyrmion. We show that the current experimental constraints give an upper bound of about $34$ TeV, which is expected to be significantly lowered near future. The impact of the existence of the EW-Skyrmion on cosmology will be discussed, and we show that constraint from the direct detection dark matter experiments give a lower bound of $O(1)$ TeV.

The paper is organized as follows. In section 2, we give a brief review of the original Skyrme model. Then, in section 3, we introduce a light scalar (higgs) into the system, and derive the coupled equations which should be satisfied by the topological object. It will be shown that a non-trivial topological configuration actually exists in the system. In section 4, upper bound on the mass of the EW-Skyrmion will be derived, and in section 5, cosmological impact of the existence of the EW-Skyrmion will be discussed. 

\section{Skyrme Model}
In order to provide a theoretical foundation and to introduce various notations, before going to the system which includes a scalar particle, we review the scaled-up version of the Skyrme model in this section. The Lagrangian is obtained by simply replacing the pion decay constant $f_\pi$ in the original Skyrme model Lagrangian by the EW scale $v_{\rm EW}$:
\begin{equation}
{\cal L} \,=\, \frac{v_{\rm EW}^2}{4}\,  {\rm Tr}\left[  \partial_\mu U(x)\,  \partial^\mu U(x)^\dagger  \right]         
            \,+\, \frac{1}{32 e^2}\, {\rm Tr}\Big[ \left[V_\mu (x) , V_\nu(x)\right] \left[V^\mu (x) , V^\nu(x)\right]  \Big],
            \label{eq:L-Skyrme}
\end{equation}
where
\begin{equation}
 U(x) = e^{i\, \pi^i(x)\, \sigma^i/v_{\rm EW} }\ \ \ \ \left(\sigma^i : {\rm Pauli\ matrix}\right),
\end{equation}
\begin{equation}
 V_\mu (x) \equiv \left( \partial_\mu U(x) \right)\,  U(x)^\dagger,
\end{equation}
and $[A, B]\equiv AB-BA $.
$\pi(x)$ is the NG field which parametrizes the coset space of $SU(2)_L \otimes SU(2)_R/SU(2)_V$. The second term of the Lagrangian (\ref{eq:L-Skyrme}) is often called the Skyrme term, and $e$ (which should not be confused with the electric charge) is the numerical parameter which defines the magnitude of the coefficient of the Skyrme term. Skyrmion is described as a static field configuration which has the following ``hedgehog'' shape:
\begin{equation}
U(x) = e^{i F(r) \sigma^i \hat{x}_i},
\label{eq:hedgehog}
\end{equation}
where 
\begin{equation}
r \equiv \sqrt{x_i x_i},\ \ \ \hat{x}_i \equiv x_i/r .
\end{equation}
$F(r)$ can be determined by requiring that the total energy of the system is minimized. For the discussion below, it is convenient to introduce a dimensionless coordinate $\tilde{r}$ which is defined as the radial coordinate $r$ normalized by the reference scale $R$:
\begin{equation}
\tilde{r} = \frac{r}{R},\ \ \ {\rm where} \ \ R \equiv \frac{1}{e\,v_{\rm EW}}.
\end{equation}
Then the energy of the static configuration can be written as
\begin{equation}
E\left[\tilde{F}(\tilde{r})\right] = 2\pi \left( \frac{v_{\rm EW}}{e}\right) \int_0^\infty d\tilde{r}\, \left[
    \left(  \tilde{r}^2 + 2 \sin^2\tilde{F}(\tilde{r}) \right) \tilde{F}'(\tilde{r})^2  +  \left(  2 \tilde{r}^2 + \sin^2\tilde{F}(\tilde{r}) \right) \frac{\sin^2\tilde{F}(\tilde{r})}{\tilde{r}^2} \right],
\label{eq:E-Skyrme}
\end{equation}
where $\tilde{F}(\tilde{r})$ is defined as
\begin{equation}
F(r) = F(\tilde{r}R) \equiv \tilde{F}(\tilde{r}).
\end{equation}
The necessary condition which minimize $E[\tilde{F}(\tilde{r})]$ can be obtained from the Euler-Lagrange equation with respect to $\tilde{F}(\tilde{r})$:
\begin{equation}
\left(  \tilde{r}^2 + 2 \sin^2\tilde{F}(\tilde{r}) \right) \tilde{F}''(\tilde{r})
\,+\, 2\, \tilde{r}\, \tilde{F}'(\tilde{r})
\,+\, \sin 2\tilde{F}(\tilde{r}) \left( \tilde{F}'(\tilde{r})^2 -1 - \frac{\sin^2\tilde{F}(\tilde{r})}{\tilde{r}^2} \right) = 0.
\label{eq:EL-Skyrme}
\end{equation}
Note that the above equation does not explicitly depend on either $e$ or $v_{\rm EW}$. Therefore $\tilde{F}(\tilde{r})$ takes the same shape for any values of $e$ and $v_{\rm EW}$, and the total energy is proportional to $v_{\rm EW}/e$ as one can see from the expression in Eq.~(\ref{eq:E-Skyrme}). Field configurations in this system are characterized by the topological winding number 
\begin{equation}
 B = - \frac{\varepsilon_{ijk}}{24 \pi^2} \int d^3x\ {\rm Tr} \Big[ V_i V_j V_k \Big],
\end{equation}
which is identified as the baryon number in the original Skyrme model. When the system has $B=n$ ($n$: integer), $\tilde{F}(\tilde{r})$ as a solution of Eq.~(\ref{eq:EL-Skyrme}) takes the following boundary values:
\begin{equation}
\tilde{F}(0) = n \pi,\ \ \ \tilde{F}(\infty) = 0.
\end{equation}
The latter is required so that the system has a finite energy. The numerical solution of $\tilde{F}(\tilde{r})$ for the case of $n=1$ is shown as the dashed curve in Fig.~\ref{fig:F-phi}. The total energy, or Skyrmion mass $M$, can be obtained by substituting this solution into Eq.~(\ref{eq:E-Skyrme}) as
\begin{equation}
 M = 72.92 \times \frac{v_{\rm EW}}{e}.
 \label{eq:72}
\end{equation}

\section{Skyrmion in the EW chiral Lagrangian with a light scalar}
In this section, we extend the scaled-up version of the Skyrme model described in the previous section so that a light scalar degree of freedom is included in the system.\footnote{See Ref.~\cite{He:2015eua} for a related work in the context of QCD.} We introduce a scalar field $h(x)$ by linearizing the non-linear field $U(x)$ in the kinetic term of the Skyrme-model Laglangian (\ref{eq:L-Skyrme}), and add the potential of $h(x)$ in the following way:
\begin{eqnarray}
{\cal L} &=&  \frac{v_{\rm EW}^2}{4} \left( 1 + \frac{h(x)}{v_{\rm EW}} \right)^2 {\rm Tr}\left[  \partial_\mu U(x)\,  \partial^\mu U(x)^\dagger  \right]      \,+\, \frac{1}{32 e^2}\, {\rm Tr}\Big[ \left[V_\mu (x) , V_\nu(x)\right] \left[V^\mu (x) , V^\nu(x)\right]  \Big]\nonumber \\
           & & + \frac{1}{2}\partial_\mu h(x)\partial^\mu h(x) - V(h(x)),
\label{eq:L}
\end{eqnarray}
where the potential of $h(x)$ is defined as follows:
\begin{equation}
V(h(x)) = \lambda v_{\rm EW}^2\, h(x)^2 +  \lambda v_{\rm EW}\, h(x)^3 + \frac{\lambda}{4}\, h(x)^4.
\label{eq:Vh}
\end{equation}
In the limit of $e \rightarrow \infty$, the Lagrangian defined above is nothing but the SM after replacing derivatives with covariant ones. The newly added Skyrme term describes anomalous quartic interactions among EW gauge bosons. Phenomenologically, we know only the upper bound of the coefficient $1/16e^2$ from the experiments of the EW gauge boson scattering. As a possible origin, it has been known that the exchange of the vector resonance effectively generates the Skyrme term with the appropriate signature to support the Skyrmions.

In general, $h(x)$ and $U(x)$ do not have to form a single linear field, thus the factor in front of the first term of the Lagrangian (\ref{eq:L}) could be any function of $h(x)/f$ with $f$ being a scale which is not necessarily related to $v_{\rm EW}$. Also a similar factor can be multiplied to the Skyrme term as well. There could be other types of higher order terms in the Lagrangian, and the form of $V(h(x))$ does not have to be limited to the one shown in Eq.~(\ref{eq:Vh}) as well. The choice of the Lagrangian above is to make the study tractable and to make the difference from the SM Lagrangian as minimal as possible: only the difference between the Lagrangian discussed in this section and that of the SM is the existence of the Skyrme term.

\begin{figure}[t]
 \begin{center}
  \includegraphics[width=100mm]{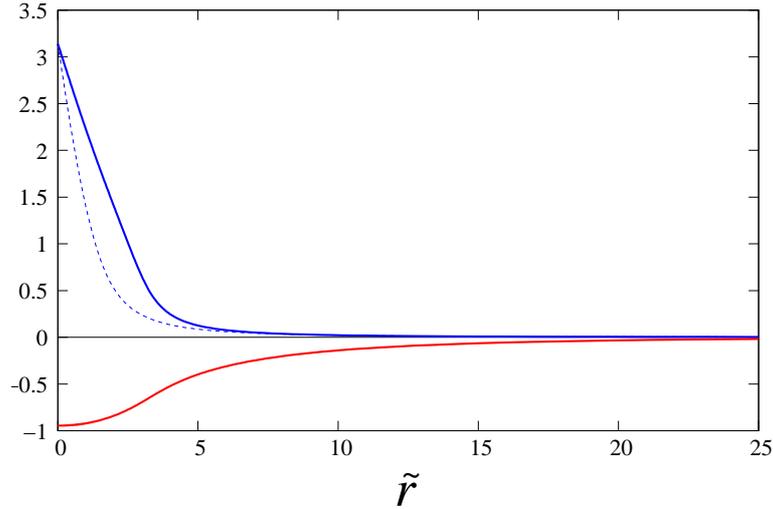}
 \end{center}
 \caption{$\tilde{F}$ (upper blue curve) and $\tilde{\phi}$ (lower red curve) as functions of $\tilde{r}$ for the case of $e = 6.5$, $v_{\rm EW} = 246$ GeV, $m_h = 125$ GeV. Dashed curve represents the $\tilde{F}$ in the case of no scalar particle.}
 \label{fig:F-phi}
\end{figure}

We take the same ansatz for the form of the static configuration of $U(x)$ as shown in Eq.~(\ref{eq:hedgehog}). As for $h(x)$, we assume that the static solution, $h_0(x)$, is spherically symmetric:
\begin{equation}
 h_0(x)/v_{\rm EW} = \phi(r).
\end{equation}
With these ansatz, the total energy of the system takes the following form:
\begin{eqnarray}
E\left[\tilde{F}(\tilde{r}), \tilde{\phi}(\tilde{r})\right] = 2\pi \left( \frac{v_{\rm EW}}{e}\right) \int_0^\infty &d\tilde{r}&\tilde{r}^2\, \Bigg[
 \  \left( 1 + \tilde{\phi}(\tilde{r}) \right)^2 \left( \tilde{F}'(\tilde{r})^2 + 2\, \frac{ \sin^2\tilde{F}(\tilde{r}) }{\tilde{r}^2} \right)
 \nonumber\\
 &&\ +\ \frac{ \sin^2\tilde{F}(\tilde{r}) }{\tilde{r}^2} \left( \frac{ \sin^2\tilde{F}(\tilde{r}) }{\tilde{r}^2} + 2 \tilde{F}'(\tilde{r})^2 \right)
 \nonumber\\
  &&\ +\  \tilde{\phi}'(\tilde{r})^2 \ +\  \frac{\,m_h^2\,}{e^2\,v_{\rm EW}^2}\left(  \tilde{\phi}(\tilde{r})^2+\tilde{\phi}(\tilde{r})^3+\frac{1}{4}\tilde{\phi}(\tilde{r})^4\right) \Bigg].\nonumber \\
\label{eq:E2}
\end{eqnarray}
Here, $\tilde{\phi}(\tilde{r})$ is defined by $\phi(r) = \phi(\tilde{r}R) \equiv \tilde{\phi}(\tilde{r})$, and we replaced the parameter $\lambda$ by the mass of the scalar $m_h$ by using the relation $\lambda=m_h^2/(2v_{\rm EW}^2)$. The energy is minimized when $\tilde{F}(\tilde{r})$ and $\tilde{\phi}(\tilde{r})$ satisfy the following coupled equations:
\begin{eqnarray}
\left(1+\tilde{\phi}(\tilde{r})\right)^2 \left( -\sin 2\tilde{F}(\tilde{r})  + 2 \tilde{r}\tilde{F}'(\tilde{r}) + \tilde{r}^2\tilde{F}''(\tilde{r})\right)
+ 2 \left(1+\tilde{\phi}(\tilde{r})\right) \tilde{\phi}'(\tilde{r})\, \tilde{r}^2\tilde{F}'(\tilde{r})  & & \nonumber\\
- \frac{ \sin^2\tilde{F}(\tilde{r}) \sin2\tilde{F}(\tilde{r}) }{\tilde{r}^2} + \sin2\tilde{F}(\tilde{r})\,\tilde{F}'(\tilde{r}) ^2  + 2\sin^2\tilde{F}(\tilde{r})\,\tilde{F}''(\tilde{r})
= &0.&\ \ 
\label{eq:Eq2}
\end{eqnarray}
\begin{eqnarray}
\left(1+\tilde{\phi}(\tilde{r})\right)\left(  \tilde{r}^2\tilde{F}'(\tilde{r}) + 2 \sin^2\tilde{F}(\tilde{r}) \right)
-2\tilde{r} \tilde{\phi}'(\tilde{r})-\tilde{r}^2 \tilde{\phi}''(\tilde{r}) && \nonumber\\
+\, \frac{1}{2}\,\frac{m_h^2}{\, e^2 \, v_{\rm EW}^2\,}\, \tilde{r}^2 \left( 2\,\tilde{\phi}(\tilde{r})  + 3\, \tilde{\phi}(\tilde{r})^2 + \tilde{\phi}(\tilde{r})^4 \right) &=& 0.
\label{eq:Eq3}
\end{eqnarray}
We numerically solved the above equations, and found that the system always has topologically non-trivial field configurations. We call this topological object as the EW-Skyrmion. In Fig.~\ref{fig:F-phi}, we show the example of $\tilde{F}(\tilde{r})$ (upper blue curve) and $\tilde{\phi}(\tilde{r})$ (lower red curve) in the case of $e=6.5$, $v_{EW}=246$ GeV, and $m_h = 125$ GeV.

\begin{figure}[t]
 \begin{center}
  \includegraphics[width=100mm, angle=0]{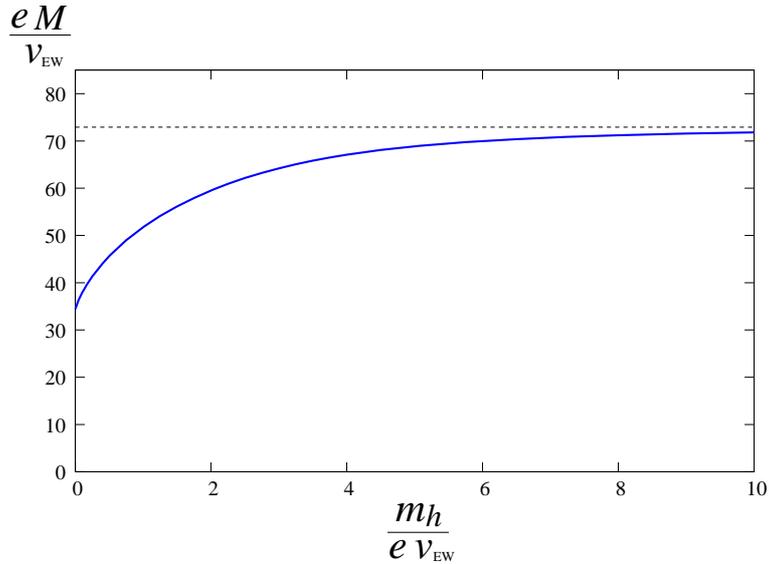}
 \end{center}
 \caption{Skyrmion mass (normalized by $\frac{v_{\rm EW}}{e}$) as a function of $\frac{m_h}{e v_{\rm EW}}$. Dashed line indicates 72.92, which is the value in the case of no scalar particle in the theory. (See Eq.~(\ref{eq:72}).)}
 \label{fig:massEWS}
\end{figure}

 As one can see from Eqs.~(\ref{eq:E2}) - (\ref{eq:Eq3}), the mass of the EW-Skyrmion normalized by $(v_{\rm EW}/e)$ depends only on the combination $m_h/(e v_{\rm EW})$. The blue curve in Fig.~\ref{fig:massEWS} shows how the mass of the EW-Skyrmion, $M$, normalized by $(v_{\rm EW}/e)$ depends on the input value of $m_h/(e v_{\rm EW})$. 
The value indicated by the dashed line in the figure corresponds to $e M/v_{\rm EW}$ in the case of no light scalar degree of freedom, namely the value derived in Eq.~(\ref{eq:72}) in the previous section. The blue curve asymptotically approaches to that value in the limit of large $m_h/(e v_{\rm EW})$. This is an expected behavior considering the fact that taking the value of $m_h/(e v_{\rm EW})$ infinity corresponds to taking the decoupling limit of the scalar particle with $e$ and $v_{\rm EW}$ being fixed. Meanwhile, when $m_h/(e v_{\rm EW})$ is taken to be smaller, the normalized mass of the EW-Skyrmion is reduced, and it approaches to about half of the value of the decoupling limit. In the next section, we use this behavior to place an upper bound on the mass of the EW-Skyrmion from currently available experimental data.

In the quantum theory of Skyrmion, the angular momentum of the configurations is quantized and the topological object is chosen to be either spin integer or half-integer, i.e., bosons or fermions. The ansatz of the shape of the solution as well as the coefficient of the Wess-Zumino-Witten (WZW) term, if exists, give relations among the topological charge, the angular momentum, and the gauge quantum number such as the electric charge \cite{Adkins:1983ya}. Hereafter, we simply choose the spin-0 and neutral solution as the lowest energy state that corresponds to the absence of the WZW term. (See Ref.~\cite{Gillioz:2010mr} for general discussion of the quantum numbers of the Skyrmion in the context of little Higgs models.)

\section{Experimental constraint}

\begin{figure}[t]
 \begin{center}
  \includegraphics[width=100mm, angle=0]{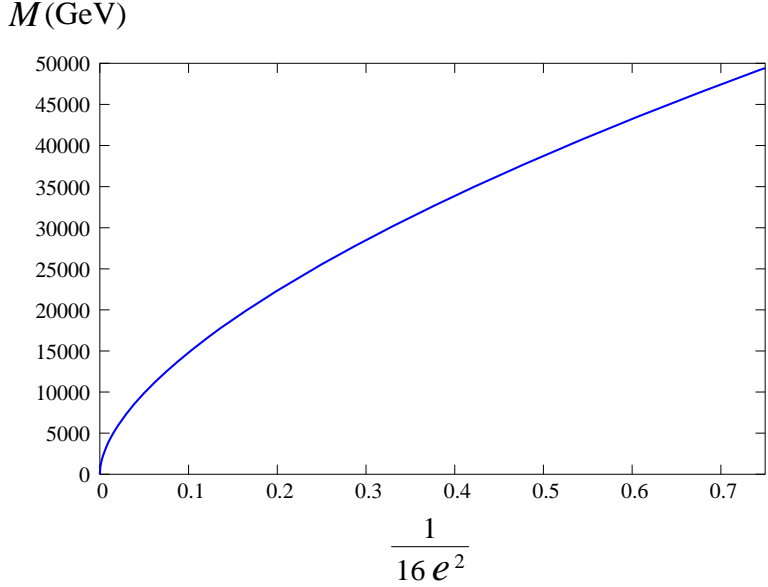}
 \end{center}
 \caption{EW-Skyrmion mass as a function of $1/(16e^2)$ with inputs $v_{\rm EW} = 246$ GeV, $m_h = 125$ GeV.}
 \label{fig:EWSmass}
\end{figure}

Among parameters in the Lagrangian (\ref{eq:L}), values of $v_{\rm EW}$ and $\lambda$, or equivalently $m_h$, are known ($v_{\rm EW} = 246$ GeV, $m_h = 125$ GeV) experimentally, while the magnitude of the coefficient of the Skyrme term have not been observed yet, and an upper bound exists through the experimental constraints on the ${\cal O}(p^4)$ terms of the EW chiral Lagrangian. The relevant terms of the EW chiral Lagrangian are conventionally expressed as~\cite{Appelquist:1993ka} 
\begin{equation}
{\cal L}_{{\cal O}(p^4)} = \alpha_4\, {\rm Tr}\left[ V_\mu V_\nu \right] {\rm Tr}\left[ V^\mu V^\nu \right]
                  \ +\        \alpha_5\, {\rm Tr}\left[ V_\mu V^\mu \right] {\rm Tr}\left[ V_\nu V^\nu \right],
\label{eq:Op4}
\end{equation}
and experimental constraints on the coefficients $\alpha_{4, 5}$ can be obtained from the scattering data of EW gauge bosons \cite{Aad:2014zda, Aad:2016ett}. The right hand side of Eq.~(\ref{eq:Op4}) can be rewritten as
\begin{equation}
-\frac{1}{2} \, \alpha_5 \,{\rm Tr}\Big[ \left[V_\mu (x) , V_\nu(x)\right] \left[V^\mu (x) , V^\nu(x)\right]  \Big]
\,+\,\frac{1}{2} \, (\alpha_4 + \alpha_5) \,{\rm Tr}\Big[ \left\{V_\mu (x) , V_\nu(x)\right\} \left\{V^\mu (x) , V^\nu(x)\right\}  \Big],
\end{equation}
where $\{A, B\} \equiv AB + BA$. The first term is nothing but the Skyrme term, while the second term is another $O(p^4)$ term, which we assumed to be absent in the current study. Therefore, to obtain the constraint on the coefficient of the Skyrme term from 95\% CL allowed region on the $\alpha_4$-$\alpha_5$ plane, one should look at the one-dimentional projection of the allowed region onto the $\alpha_4+\alpha_5 =0$. By reading the constraint on $\alpha_5$ on the line of $\alpha_5=-\alpha_4$, both Ref.~\cite{Aad:2014zda} and Ref.~\cite{Aad:2016ett} give a similar lower bound, $\alpha_5 \gtrsim\ -0.4$, which can be  translated to the upper bound on the coefficient of the Skyrme term as follows:
\begin{equation}
 \frac{1}{16 e^2}\ \lesssim\ 0.4. 
 \label{eq:constraint}
\end{equation}
In Fig.~\ref{fig:EWSmass}, we plot the mass of the EW-Skyrmion as a function of $1/(16e^2)$ with $v_{\rm EW} = 246$ GeV and $m_h = 125$ GeV being fixed. 
This figure, together with the constraint in Eq.~(\ref{eq:constraint}), give an upper bound of the mass of the EW-Skyrmion $M \lesssim 34$ TeV. Constraints on $\alpha_{4, 5}$ are expected to be significantly improved by LHC RUN2 (see, for example, Ref.~\cite{Fabbrichesi:2015hsa}), and the upper bound on the coefficient of the Skyrme term will be reduced to the level of $0.01$, which corresponds to the upper bound on the mass of the EW-Skyrmion $M \lesssim 4$~TeV.

\section{Cosmology}
In this section, we discuss observational constraints on the EW-Skyrmion as a dark matter.
We first discuss the possibility of the thermal production of the EW-Skyrmion, in which case, the relic density $\Omega_S$ is related to the pair annihilation cross section $\sigma_A$ as follows:
\begin{equation}
\Omega_S h^2 \approx \frac{3 \times 10^{-27} {\rm cm}^3/{\rm sec}}{\langle \sigma_A v_{\rm rel}\rangle},
\label{eq:relic}
\end{equation}
Here, we assume that the geometric cross section ($\sim \pi R^2$, where $R=1/(e\, v_{\rm EW})$ is a typical radius of the EW-Skyrmion) is a reasonable estimate of the annihilation cross section, and substitute $\langle \sigma_A v_{\rm rel}\rangle = \pi/(e\, v_{\rm EW})^2$ to obtain a rough estimate of the right hand side of Eq.~(\ref{eq:relic}). By requiring that the relic density of the EW-Skyrmion does not exceed the current dark matter density, $\Omega_{\rm DM}h^2\approx 0.1$, we obtain the following constraint on the parameter $e$:
\begin{equation}
 e \lesssim 150.
\end{equation}
When $e$ takes the maximum value, $e \simeq150$, in which case the relic density of the dark matter is explained solely by the EW-Skyrmion, the mass of the EW-Skyrmion is about $M = 60$ GeV. The EW-Skyrmion with such a small mass is excluded by the direct detection experiment of the dark matter, such as the LUX experiment \cite{Akerib:2015rjg}, unless the coupling relevant to the direct detection is extremely small, which is unlikely in the case of the current scenario. If the mass of the EW-Skyrmion is larger, the constraint from the direct detection experiments become weaker,  however, in that case, the amount of the contribution of the EW-Skyrmion to the current dark matter density becomes smaller. Also, when the mass of the EW-Skyrmion is too large, say $M= 10$ TeV, the freeze-out temperature of the EW Skyrmion becomes much higher than the EW scale. Since the EW-Skyrmion exists only in the broken phase of the EW symmetry, it is not self-consistent to discuss the EW-Skyrmion beeing in the thermal equilibrium at the temperature much higher than the critical temperature of the EW phase transition. 

From the above considerations, it is unlikely that the thermal relic component of the EW-Skyrmion becomes the dominant source of the dark matter of the current Universe. Since the EW-Skyrmion has the conserved quantum number (topological winding number), the asymmetry of the quantum number in the full description of the theory may have been produced in the history of the Universe \cite{Nussinov:1985xr} and that can remain today as the energy density carried by the EW-Skyrmion. The mechanism how such a production occurs depends on the UV physics which realizes the low-energy effective picture of the EW-Skyrmion as well as the history of the inflationally Universe. Instead of trying to specify those, we discuss the consistency of the asymmetry scenario with existing data and future experimental prospects. 

In order to discuss the constraint from the direct detection dark matter experiments, we need to know how the EW-Skyrmion interacts with SM particles, especially with the Higgs. Calculating couplings requires the precise definition of the UV physics and the proper treatment of the extended object. Here, instead of trying to be precise, we adopt a simple assumption that the EW-Skyrmion has a low-energy effective coupling to the Higgs doublet with the form ${\cal L}_{\rm eff} = -2\kappa |S|^2 |H^2|$, where $H$ is the Higgs doublet and $S$ is the EW-Skyrmion field. (See, Ref.~\cite{Silveira:1985rk} for analyses of the equivalent effective model.) The purpose here is to give a rough idea of the cosmological impact of the EW-Skyrmion. With this assumption, the spin-independent elastic EW-Skyrmion-nucleon cross section as a function of the EW-Skyrmion mass $M$ can be expressed as follows:
\begin{eqnarray}
\sigma_{\rm SI} &\approx& \frac{\kappa^2 m_N^4 f^2}{\pi M^2 m_h^4} \\
 &\simeq&  \left( \frac{\kappa}{1.0} \right)^2 \left(\frac{1\, {\rm TeV}}{M} \right)^2 \left(\frac{f}{0.3} \right)^2 \times 3.6 \times 10^{-44}\ {\rm cm}^2,
\end{eqnarray}
where $m_N$ is the nucleon mass and $f$ is the form factor, which is taken to be $0.3$ here. The upper bound on the WIMP-nucleon cross section given by the LUX experiment \cite{Akerib:2015rjg} can be applied to the above cross section, and we obtain $M \gtrsim 1.5$ TeV when we assume $\kappa=1.0$. Taking larger (smaller) value of $\kappa$ places stronger (weaker) constraint: when the coupling is taken to be $\kappa = \pi\, (0.5)$, the bound becomes $M\gtrsim 3.5$ TeV (1 TeV). Combining this lower bound with the upper bound obtained in the previous section, we conclude that the EW-Skyrmion with mass between $1.5$ TeV and $34$ TeV, with certain amount of uncertainty, is consistent with current experimental bounds, while explaining the dark matter of the Universe. As was mentioned in the previous section, LHC RUN2 will give stronger upper bound, $M\lesssim 4$ TeV, which will probe the most of the allowed region obtained here. The sensitivity of the direct detection dark matter experiments will be also improved, therefore, there is a possibility that we will find a hint of the dark matter in near future.

\section{Summary}
In this paper, we formulated coupled equations for a topologically stable object in the system of the electroweak chiral Lagrangian with a light scalar boson. We applied those to the Higgs sector of the Standard Model with minimal addition of the $O(p^4)$ (Skyrme) term, and showed that a non-trivial solution exists in the system. From the experimental constraint on the magnitude of the coefficient of the Skyrme term, upper bound on the mass of the topological object  (EW-Skyrmion) was derived as $M\lesssim 34$ TeV. We discussed observational constraint on the EW-Skyrmion as a dark matter, and showed that non-thermally produced EW-Skyrmion which is heavier than $O(1)$ TeV is consistent with the direct detection dark matter experiments. Both upper and lower bounds derived in this paper will be improved by LHC RUN2 and future direct detection dark matter experiments, and currently allowed mass region of the EW-Skyrmion will be entirely probed near future.

\acknowledgments
This work was supported by JSPS KAKENHI Grant-in-Aid for Scientific Research (B) (No. 15H03669 [RK]) and MEXT KAKENHI Grant-in-Aid for Scientific Research on Innovative Areas (No. 25105011 [RK,MK]). We would like to thank Ryosuke Sato for discussions.


\begin{thebibliography}{99}

\bibitem{Skyrme:1961vq} 
  T.~H.~R.~Skyrme,
``A Nonlinear field theory,''
  Proc.\ Roy.\ Soc.\ Lond.\ A {\bf 260}, 127 (1961).
  doi:10.1098/rspa.1961.0018

\bibitem{Ellis:2012bz} 
  J.~Ellis and M.~Karliner,
``Indications on the Mass of the Lightest Electroweak Baryon,''
  Phys.\ Lett.\ B {\bf 713}, 233 (2012)
  [arXiv:1204.6642 [hep-ph]].
  
\bibitem{Ellis:2012cs} 
  J.~Ellis, M.~Karliner and M.~Praszalowicz,
``Generalized Skyrmions in QCD and the Electroweak Sector,''
  JHEP {\bf 1303}, 163 (2013)
  [arXiv:1209.6430 [hep-ph]].
    
\bibitem{He:2015eua} 
  B.~R.~He, Y.~L.~Ma and M.~Harada,
  ``Effects of scalar mesons in a Skyrme model with hidden local symmetry,''
  Phys.\ Rev.\ D {\bf 92}, no. 7, 076007 (2015)
  doi:10.1103/PhysRevD.92.076007
  [arXiv:1507.00437 [hep-ph]].
   
\bibitem{Adkins:1983ya} 
  G.~S.~Adkins, C.~R.~Nappi and E.~Witten,
 ``Static Properties of Nucleons in the Skyrme Model,''
  Nucl.\ Phys.\ B {\bf 228}, 552 (1983).
  doi:10.1016/0550-3213(83)90559-X

\bibitem{Gillioz:2010mr} 
  M.~Gillioz, A.~von Manteuffel, P.~Schwaller and D.~Wyler,
  ``The Little Skyrmion: New Dark Matter for Little Higgs Models,''
  JHEP {\bf 1103}, 048 (2011)
  doi:10.1007/JHEP03(2011)048
  [arXiv:1012.5288 [hep-ph]],
  M.~Gillioz,
  ``Dangerous Skyrmions in Little Higgs Models,''
  JHEP {\bf 1202}, 121 (2012)
  Erratum: [JHEP {\bf 1303}, 123 (2013)]
  doi:10.1007/JHEP02(2012)121, 10.1007/JHEP03(2013)123
  [arXiv:1111.2047 [hep-ph]].
   
\bibitem{Appelquist:1993ka} 
  T.~Appelquist and G.~H.~Wu,
  ``The Electroweak chiral Lagrangian and new precision measurements,''
  Phys.\ Rev.\ D {\bf 48}, 3235 (1993)
  [hep-ph/9304240].

\bibitem{Aad:2014zda} 
  G.~Aad {\it et al.} [ATLAS Collaboration],
  ``Evidence for Electroweak Production of $W^{\pm}W^{\pm}jj$ in $pp$ Collisions at $\sqrt{s}=8$ TeV with the ATLAS Detector,''
  Phys.\ Rev.\ Lett.\  {\bf 113}, no. 14, 141803 (2014)
  doi:10.1103/PhysRevLett.113.141803
  [arXiv:1405.6241 [hep-ex]].
  
\bibitem{Aad:2016ett} 
  G.~Aad {\it et al.} [ATLAS Collaboration],
  ``Measurements of $W^\pm Z$ production cross sections in $pp$ collisions at $\sqrt{s} = 8$ TeV with the ATLAS detector and limits on anomalous gauge boson self-couplings,''
  arXiv:1603.02151 [hep-ex].

\bibitem{Fabbrichesi:2015hsa} 
  M.~Fabbrichesi, M.~Pinamonti, A.~Tonero and A.~Urbano,
  ``Vector boson scattering at the LHC: A study of the WW $\to$ WW channels with the Warsaw cut,''
  Phys.\ Rev.\ D {\bf 93}, no. 1, 015004 (2016)
  doi:10.1103/PhysRevD.93.015004
  [arXiv:1509.06378 [hep-ph]].
  
\bibitem{Akerib:2015rjg} 
  D.~S.~Akerib {\it et al.} [LUX Collaboration],
  ``Improved Limits on Scattering of Weakly Interacting Massive Particles from Reanalysis of 2013 LUX Data,''
  Phys.\ Rev.\ Lett.\  {\bf 116}, no. 16, 161301 (2016)
  doi:10.1103/PhysRevLett.116.161301
  [arXiv:1512.03506 [astro-ph.CO]].
  
\bibitem{Nussinov:1985xr} 
  S.~Nussinov,
  ``Technocosmology: Could A Technibaryon Excess Provide A 'natural' Missing Mass Candidate?,''
  Phys.\ Lett.\ B {\bf 165}, 55 (1985).
  doi:10.1016/0370-2693(85)90689-6
  
\bibitem{Silveira:1985rk} 
  V.~Silveira and A.~Zee,
  Phys.\ Lett.\ B {\bf 161}, 136 (1985).
  doi:10.1016/0370-2693(85)90624-0
    
\end{thebibliography}
\end{document}